\documentclass[conference]{IEEEtran}
\IEEEoverridecommandlockouts
\usepackage{cite}
\usepackage{amsmath,amssymb,amsfonts}
\usepackage{algorithmic}
\usepackage{graphicx}
\usepackage{textcomp}
\usepackage{xcolor}

\def\BibTeX{{\rm B\kern-.05em{\sc i\kern-.025em b}\kern-.08em
    T\kern-.1667em\lower.7ex\hbox{E}\kern-.125emX}}
\newboolean{showcomments}
\setboolean{showcomments}{true} 
\ifthenelse{\boolean{showcomments}}
  {\newcommand{\nb}[2]{
    \fcolorbox{gray}{yellow}{\bfseries\sffamily\scriptsize#1}
    {\sf\small$\blacktriangleright$\textit{#2}$\blacktriangleleft$}
   }
   
  }
  {\newcommand{\nb}[2]{}
   
  }

\begin{document}

\title{Software Architecture for ML-based Systems: What Exists and What Lies Ahead}

\author{\IEEEauthorblockN{Henry Muccini}
\IEEEauthorblockA{
\textit{University of L'Aquila}\\
L'Aquila, Italy \\
henry.muccini@univaq.it}
\and
\IEEEauthorblockN{Karthik Vaidhyanathan}
\IEEEauthorblockA{
\textit{Gran Sasso Science Institute \&
University of L'Aquila}\\
L'Aquila, Italy \\
karthik.vaidhyanathan@univaq.it}}

\maketitle

\begin{abstract}
The increasing usage of machine learning (ML) coupled with the software architectural challenges of the modern era has resulted in two broad research areas: i) software architecture for ML-based systems, which focuses on developing architectural techniques for better developing ML-based software systems, and ii) ML for software architectures, which focuses on developing ML techniques to better architect traditional software systems. In this work, we focus on the former side of the spectrum with a goal to highlight the different architecting practices that exist in the current scenario for architecting ML-based software systems. We identify four key areas of software architecture that need the attention of both the ML and software practitioners to better define a standard set of practices for architecting ML-based software systems. We base these areas in light of our experience in architecting an ML-based software system for solving queuing challenges in one of the largest museums in Italy.

\end{abstract}

\begin{IEEEkeywords}
Software Architecture, Machine Learning, Artificial Intelligence, Architecture Framework, Architecting Process, Self-adaptive Architecting, Architecture Evolution
\end{IEEEkeywords}

\section{Introduction}
Modern software systems generate a tremendous amount of data. In fact, we live in a data-driven world powered by software where we have an abundance of data generated by different sources like web applications, smartphones, sensors, etc~\cite{analyticsage,datadrivense}. A recent article from Forbes quotes that about 2.5 quintillion bytes of data are created every day.
This number is expected to increase drastically in the years to come~\cite{dataworld}.
Over the years, with the advancements in computing infrastructure, these data have been fueled by Artificial Intelligence (AI), in particular, Machine Learning (ML), to generate actionable insights~\cite{mlalpadyn}. It has further paved the way for developing software systems and services that power autonomous vehicles, recommendations in Netflix, search results in Google, etc. As remarked by Andrew Ng, AI is considered the new electricity and is expected to transform the world just like electricity did about 100 years ago~\cite{ai_elec}. Moreover, as per Gartner, around 40\% of the world's organizations are expected to leverage AI in the coming years~\cite{gartnerai}. 


However, the increasing adoption of AI, particularly ML, has given rise to different challenges associated with development practices, deployments, ensuring data quality, etc., in addition to the challenges of a traditional software system. These challenges call for better architecting practices for addressing the concerns of ML-based software systems~\cite{sa_ai_ivica,sa_ai_murphy}.
On the one hand, we have software systems that generate a tremendous amount of data but face different architectural challenges. Some of those challenges can be solved using ML~\cite{sa_ai_antony,nemi2019}, and on the other hand, we have ML-based systems that thrive on data but require better architecting practices. 
This combination of challenges in the field of software architecture and ML has resulted in two broad research areas: i) Software architecture for ML systems. It primarily focuses on developing architectural techniques for better developing ML systems;  ii) ML for Software architectures, which focuses on developing ML techniques to better architect software systems. In this work, we focus on the former side of the spectrum, intending to provide an overview of the different architecting practices prevalent in the current scenario when building ML-based software systems. We focus on four key areas of software architecture and highlight what lies ahead in each of those areas. We base our discussions in the light of our experience in building an ML-based software system for solving queuing issues in one of the largest museums in Italy. We believe the discussions put forward in this work simulate more extensive discussion among the ML and software architecture research/practitioner community members to better define practices for architecting ML-based software systems. 


\section{Software Architecture and Machine Learning}

Ever since Perry and Wolf~\cite{safoundation} came up with the seminal work on software architecture, the field has evolved continuously to cope with the different computing trends. The role of software architect has also evolved over the years based on the various advancements in domains (automotive, robotics, IoT, etc.), the practices/methodologies (waterfall, RUP, Agile, etc.), and type of applications (monolithic, SOA, microservices, etc.). On the other hand, the field of machine learning has advanced rapidly with the availability of a larger amount of data, better computing infrastructure supplemented by the increasing number of domain experts. This revolution has spanned across the different software domains, and as the years progress, software systems, by leveraging ML, are moving from the notion of being another software to more intelligent software systems. This implies we need architecting practices that allow modern software architects to handle the nuances of two different continuously changing worlds. On one side, it is about how to architect a software system that satisfies all the functional and non-functional requirements. On the other side, it is about how to handle the different challenges associated with machine learning. However, even though they co-exist in one software in the current world, there is a clear division in terms of the process, stakeholders, concerns, and so on.

\begin{figure}[h]
 \centering
 \includegraphics[width=\linewidth]{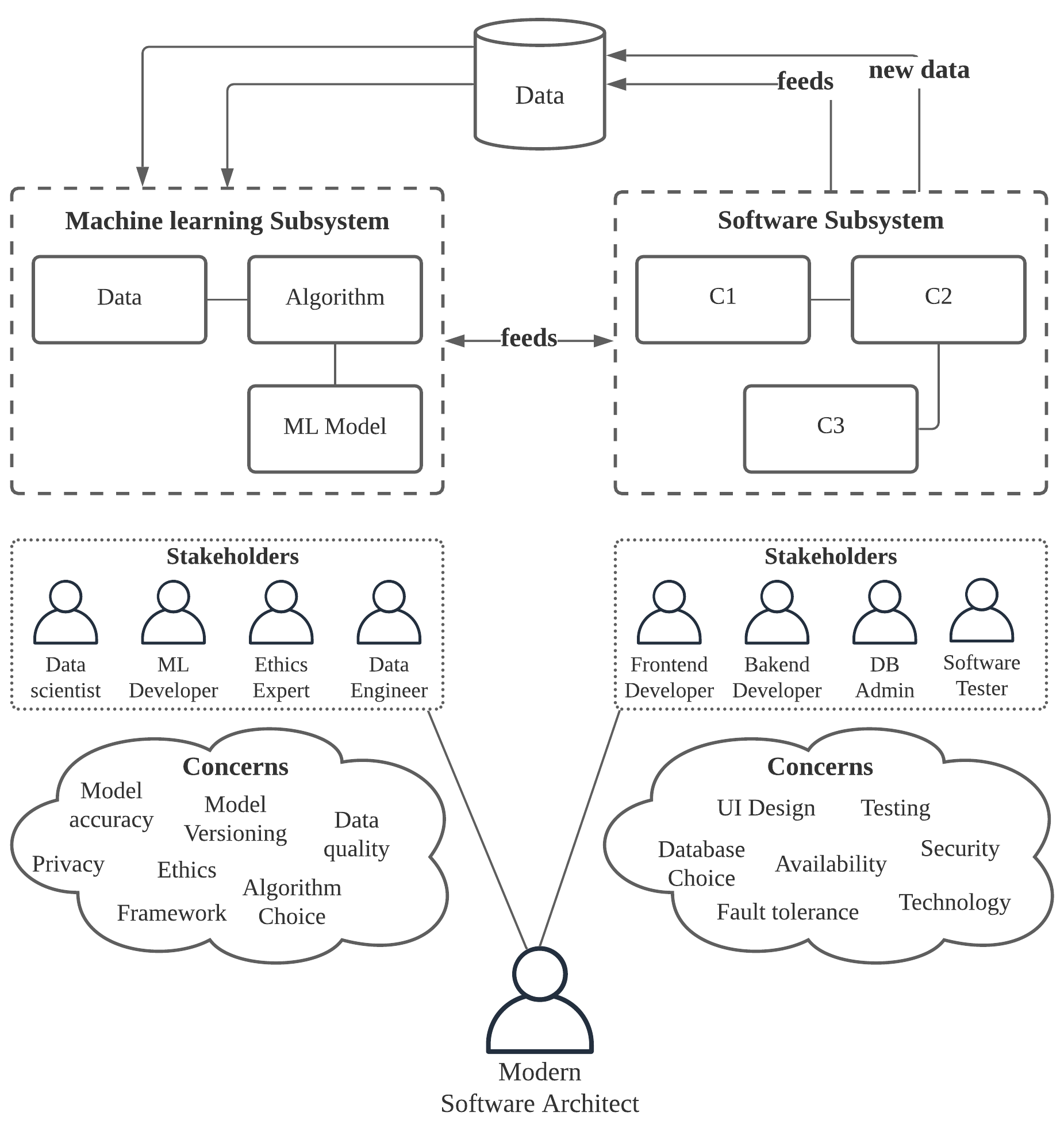}
 \caption{High-level view of an ML-based Software System}
  \label{fig:saml_modern}
\end{figure}

\noindent Figure \ref{fig:saml_modern} shows the high-level view of an ML-based software system and how a modern software architect perceives such a system. Even though it is one system, they exist as two big subsystems where on one side (machine learning subsystem), it is more about data, algorithms, and models. In contrast, on the other side (software subsystem), it is about components, connectors, and the interaction between them. The software subsystem makes use of the machine learning models and continuously produces the data required for the machine learning subsystem. Further, for each of the subsystems, there are different stakeholders and their respective concerns. It is then the role of the modern software architect to coordinate between the two different subsystems with totally different characteristics, properties, and team dynamics. This process, however, raises a lot of questions on the different aspects of architecting, ranging from how the architecting practice can be standardized to how this setup has resulted in different barriers in the process of software architecting. In the remainder of the paper, we describe the existing practices in architecting ML-based software systems and, we further elicit what lies ahead for the community to develop better approaches to architect ML-based software systems. Based on the discussions made, we also provide some pointers on how the role of a future software architect may look like.

\section{Architecting ML-based Software Systems: The Case of Uffizi Gallery}
\label{sec:uffizi}
The Uffizi Gallery in Florence is one of the most visited museums in Italy. In 2018, it counted more than 2 million visitors. Other than for its priceless masterpieces from Botticelli, Michelangelo, Raffaelo, Leonardo, and many more, Uffizi is also known for potentially long waiting queues. The reasons are manifold: i) as reported by the 2016 ISTAT Report on Tourism~\cite{itstat}, 20 museums over a total of 4,976 in Italy, attract one-third of all visitors (Uffizi being one of those); ii) Uffizi has a limited capacity that, in some periods of the year, is much smaller than the number of visitors arriving at any time; iii) the museum does not bound the time required for a visit, and therefore tourists may stay as long as they want.

Based on a joint project between the Uffizi Gallery and the University of L'Aquila, started in 2016, our team was made responsible for solving the queue problem in Uffizi.

Visitors' flow management in famous and large museums involves different stakeholders.  In the Uffizi case, we had to collect information from the Director of the Uffizi Gallery, from the Concessionaire responsible for the ticketing and welcome services, from the external Tour Operators and, finally, from visitors themselves. We realized that many reasons could cause queues in the Gallery. Among the most relevant are the following: A large number of tourists come for just one day visit to Florence and the Uffizi Gallery. These numbers are very high in peak season and on certain days of the week. However, the museum has a limited capacity and cannot accommodate more than a fixed number of people at the same time. As soon as this number is reached, no one can enter the museum until someone leaves. When the museum gets crowded and the number of people wishing to enter outnumbers the people leaving, the waiting queue starts to grow. Two factors may exacerbate this phenomenon: i) the tourists visit is not time-bounded (therefore, if they decide to stay longer, the waiting queue tends to increase), ii) there exist two separate queues: one priority line for visitors with a reservation, and a low priority queue for those without a reservation. When the priority line grows, the unbooked one may go into starvation. We considered all these different factors, along with external factors like weather, nature of the season, etc., to come up with an ML-based ticketing software system that was able to reduce the waiting time from hours to minutes~\cite{telegraph}.

We followed a typical data-driven process to realize the solution. The overall solution was a system that used IoT (with standard data pipeline architecture for data ingestion, monitoring, and storage), machine learning, statistical optimization, and data analytics. Such a solution was accomplished with the help of a team comprising of members of varying skills, including software developers, statisticians, machine learning experts, hardware engineers, and IoT experts.

While our experience is specific to smart tourism and the Uffizi Gallery, we believe that the same can be generalized to any ML-based software system. It also represents the needed intertwine between data acquisition with ambient intelligence, IoT, predictive and prescriptive models, and front-end solutions. The applied process is highly iterative and incremental, with new questions arising from time to time and new solutions being proposed incrementally. During the process, there were different challenges involved in architecting the overall solution. The main reason was that there were no standard practices to be followed to architect this kind of system where the dynamics are entirely different from a traditional software system. In the following sections, we base our discussion on the lessons learned, and the experience gathered while architecting the overall system.


\section{Software Architecture For ML-based Systems: What we miss?}
This section discusses four key areas that require the attention of both the research and practitioner community to standardize the architecting practices for ML-based software systems. We provide details on what exists in these different areas and what we believe needs to be done in the future. These areas have been derived based on our expertise in the field of software architecture and the experience in architecting ML-based software system for queue management in the Uffizi gallery (refer Section \ref{sec:uffizi}).







\subsection{Architecture Framework}
An architecture framework (AF) is a coordinated set of viewpoints, conventions, principles, and practices for architecture description within a specific domain of application or community of stakeholders~\cite{ieee42010}. They are knowledge structures that enable software architects to organize the architecture description into complementary views based on the application domain. 

\smallskip
\noindent \textbf{What exists:} Different architectural frameworks have evolved from the time the foundations of software architecture were laid by Perry and Wolf~\cite{safoundation}. These include but not limited to: \textit{Zachman Framework} by Zachman~\cite{zachman},~\textit{4+1 views} by Kruchten~\cite{kruchten},~RM-ODP Framework by Raymond~\cite{raymond},~etc. They allow expressing the system's architecture through different viewpoints by considering the different stakeholders and their concerns. However, as pointed out by Eoin Woods \cite{woods}, these do not provide support for data or algorithms, which are fundamental building blocks of ML-based systems as most of the time, the availability of data determines the choice of the ML technique to be used. Further, they also do not take into consideration the different stakeholders and concerns that arise with ML-based software system (as depicted in Figure \ref{fig:saml_modern}).

\smallskip
\noindent \textbf{What lies ahead:} As shown in Figure~\ref{fig:saml_modern}, the domain of ML, in general, brings in many new stakeholders and concerns which cannot be generalized to the standard software architecture practices. These cannot be captured using the existing architectural frameworks and requires the development of novel architectural framework(s) that shall consider the following:

\smallskip
\noindent \textit{1. Stakeholders}: 
The key stakeholders of a traditional software system consists of mainly software developers, testers, database admin, etc. (stakeholders of software subsystem in Figure \ref{fig:saml_modern}). However, unlike traditional software systems, key stakeholders are mostly from a mathematical/statistical background in an ML-based software system. These include but are not limited to Data Scientists,  ML Developers, Ethics experts, Data Engineers, etc.~(stakeholders of machine learning subsystem in Figure \ref{fig:saml_modern}). Further, they can also include members from the operations team as well as software developers who develop components that consume the ML models.

\smallskip
\noindent \textit{2. Concerns}: Each stakeholder may have one or more concerns that the system needs to address/satisfy. While the concerns of stakeholders of a traditional software system deal with choice of technologies, testing, fault tolerance, availability, security, etc., ML-based software system with an additional set of stakeholders brings in different concerns. These include data quality, ethics, data privacy, fairness, trustworthiness, ML models' performance, etc.

\smallskip
\noindent \textit{3. Viewpoints}: 
An architectural viewpoint is a collection of patterns, templates, and conventions for creating a view~\cite{ieee42010}. Such a view can be used to represent the concerns of one or more stakeholders of the system. In an ML-based software system, many involved stakeholders with varying skill sets require the architect to use multiple viewpoints to capture the different concerns. For instance, it might be required to create a {\em learning viewpoint} to support the construction of a view to represent the machine learning flow, machine learning components, etc. This view can then capture the concerns of the different stakeholders like ML developers, data scientists, ethics experts, etc. Further, a {\em data viewpoint} might be required to build a view to represent the data components, model the data pipeline, etc., that capture the concerns of stakeholders like data engineers, data scientists, etc.


\noindent These different stakeholders, concerns, and viewpoints in ML-based software systems give rise to three key potential open research areas that require the attention of members from both the software architecture as well as the ML community.

\smallskip
\noindent i) How to define an AF for machine learning systems? Such a framework can provide means to establish a common architecting standard that needs to be followed while architecting ML-based software systems or any intelligent systems in general.

\smallskip
\noindent ii) The software system will itself be developed following an AF based on the application domain. Since the software system itself may obey its AF, we expect that the ML-based software system will comprise the interaction between multiple AFs. This opens up a new research area which should focus on how correspondences rule(s) can be established between the AFs used for architecting the ML subsystem and the one used for architecting the software subsystem.

\smallskip
\noindent iii) It would also be interesting to study and define if, especially in legacy software systems where machine learning needs to be incorporated a posteriori, is it then enough to just have multiple viewpoints to define the machine learning system and then establish correspondence rules with the already existing viewpoints of the system?


\smallskip
\noindent \textbf{Experience:} As explained in Section \ref{sec:uffizi}, a team composed of experts in Software Engineering, Operational Research, Statistics, and IoT had to work on the UFFIZI project. The performance was a primary concern we had to handle carefully since the system released up to 850 tickets in a time frame of just 15 minutes. Security was another concern since the denial of service could essentially block the release of the tickets while having the museum only partially crowded. Ethics was also a concern due to the (potential) need to use personal info, such as fingerprint, face model, etc., to minimize secondary ticketing. While we had a framework, CAPS\cite{CAPS}, to model the software architecture, IoT, and digital space dimensions, we did not have a viewpoint for the ML concern.

\subsection{Architecting Process}
An architecting process deals with the different activities and tasks involved in the production of architectural solutions for software systems. 

\noindent \textbf{What exists:} There have been works that report on the interplay between software engineering and machine learning workflow, such as the one presented by Amershi et al.~\cite{amershi2019}. The work also discusses in detail how these interplay in workflow results in a constant back and forth communication between software development and machine learning teams. This process requires the software development teams to be upskilled with more in-depth knowledge of ML to enable smoother interaction between such teams.
Further, the impact of using machine learning on different software development practices starting from requirements gathering to testing and maintenance was discussed by Zhiyuan et al.~\cite{sa_ai_murphy} while a taxonomy of software engineering challenges for machine learning systems was discussed by Lwakatare et al. \cite{lwakatare2019taxonomy}. However, these work focus on the engineering aspects and are not specialized in architectural artifacts and solutions.

\smallskip
Zhiyuan et al. state that ML development is an incrementally iterative process~\cite{sa_ai_murphy} where many times the development starts with performing experiments to understand what algorithm/technique suits the best to satisfy a particular requirement. This may imply modifying the architecture of the system to meet the needs of the algorithm/technique (e.g., through better mechanisms to collect data or additional components for pre-processing). This type of development would also mean that there is a constant back and forth interaction between software architects and stakeholders of the ML components (who speak more of a mathematical language) as well as between the software architects and the stakeholders of the software system (who speak a programmatic language). The nature of two different development methodologies and two different teams with totally different backgrounds can impact the various stages of the architecting process\footnote{The different process are based on the different stages prescribed by the International Federation of Information Processing Working Group 2.10}:

\smallskip
\noindent \textit{1. Architecture Design:} In an ML-based software system, due to the existence of two different subsystems (as depicted in Figure \ref{fig:saml_modern}), the first potential area that needs to be investigated is understanding what drives the architectural design of an ML-based software system. Is it the concerns of the ML subsystem or the concerns of the software subsystem? It is more like an egg or chicken problem. One possible way to handle this is that the software architect starts from a more significant design concern, breaks it down into sub concerns, and then maps them into design decisions within each subsystem. The second broad area is how to manage trade-offs between cross-cutting design decisions that span across both subsystems. This is because on one side, in the ML subsystem, design decisions will be more driven by data and algorithm requirements, whereas, on the other side, it will be more based on the software functional and non-functional requirements. So how to balance trade-offs in scenarios where achieving non-functional requirements requires the degradation of ML performance or vice versa.

\smallskip
\noindent \textit{2. Architecture Analysis:} One of the vital steps in the process of architecting after the design is to analyze the architecture. Such an analysis is accomplished in a traditional software system by generating informal, formal, or mathematical models with queuing networks, quantitative verification, etc.{~\cite{archanalysissurvey,quantitativeverification}}. These are then used to determine if the system satisfies specific quality attributes. However, in an ML-based software system, the very nature of the process itself is {\em probabilistic} in nature. This is because the output of an ML algorithm itself is based on a probabilistic measure. Hence, quantifying the flow on components, performing analysis, and verifying different functional and non-functional properties in such a system requires sophisticated mechanisms.

\smallskip
\noindent \textit{3. Architecture Realization:} Most software development these days follows an agile methodology. On the contrary, ML development follows an increasingly iterative process. Moreover, agile follows the principle of {\em user first}, where the development is based on user stories. In contrast, in ML-based software systems, the development is based on {\em features} that may or may not directly impact the user. Hence the development team might use two different methodologies for development, which needs to be better managed. The organizational aspect of such teams also needs to be considered. There might be people with different expertise, so the need for a {\em central coordinator} who understands both the software and ML world is increasingly becoming important.
 
\smallskip
\noindent \textit{4. Architecture Representation:} One of the expected results of any software architecting process is to generate artifacts that can be shared with the concerned stakeholders to better communicate the system's architecture. In a traditional software system, this is achieved with the help of different types of UML diagrams or even simple box and lines/arrow diagrams. However, in the case of ML-based software systems, this may not be possible. The main reason for this can be attributed to the fact that ML requires a lot of mathematical understanding, and communicating these to the business stakeholder(s) is often challenging (as artifacts are often produced models, versioned datasets, etc.) This brings back to the interplay between the {\em introvert} nature of an architect (with skills on formal reasoning and analysis) and her {\em extrovert} nature, highlighting her role as communicator and negotiator\cite{6374194}.

\smallskip
Establishing a standard set of processes for architecting ML-based systems is fundamental given the ever-growing usage of ML in production. This implies we need extensive research to be done to:

\smallskip
\noindent i) understand how to better manage and document {\em design decisions} which also considers the data quality, data volume, probabilistic nature of the learning algorithms, etc., as well as the constraints on software systems. 

\smallskip
\noindent ii) develop methodologies with tool support that allows architects to {\em analyze} the architecture of ML-based software systems.

\smallskip
\noindent iii) establish a set of practices that need to be followed to better organize the development teams such that the software development team and ML team can work in a more cohesive environment, thereby leading to smoother functioning of teams.

\smallskip
\noindent iv) develop tools and techniques that can provide rich visualizations of the data and models generated behind the scenes. This will also enable the architects to communicate better the effort required in realizing such complex systems.

\smallskip
\noindent \textbf{Experience:} In the Uffizi project, there was a continuum of data selection, ML model creation, and architecture realization. There was not a systematic process, but the continuous interleave between ML-based decisions (i.e., which data to generate, which data to analyze and store, which ML model to use) and software/hardware-based decisions (i.e., how to collect the data in a historical place, how to maximize performance, how to maximize precision). This imposed the linearization of many tasks, thus reducing development speed. We adverted the need for a more systematic process to clearly and effectively divide the workload among team members. Moreover, the software engineering team had to learn the data-scientist language, the ML expert jargon, and vice-versa. 


    

\subsection{Self-Adaptive Architecting}
Self-adaptative architecture equips the system with the ability to autonomously adapt the architecture of the system (with minimal human intervention) to handle the different types of uncertainties, which may lead to Quality of Service degradation and failure.

\smallskip
\noindent \textbf{What exists:} The field of \textit{Self-adaptive systems} evolved over the years from the term \textit{"Software Crisis"} which was first coined in 1968 at the NATO Software Engineering Conference~\cite{nato}. Modern software systems are subjected to various types of uncertainties due to resource constraints, component failures, etc. Towards this over the years, self-adaptation continues to emerge as a potential solution
~\cite{weyns2020introduction} as it enables the architectures to autonomously adapt to the different uncertainties so as to satisfy the overall goals of the system. There has been an extensive amount of literature work that has been done in the area of self-adaptive systems. An elaborate survey on different approaches to engineer self-adaptive systems was presented by Macías-Escrivá et al.~\cite{sastwo}, and Krupitzer et al.~\cite{sassurvey}. In recent years, there have been different works done in the self-adaptation literature that makes use of machine learning techniques to better adapt traditional software systems~\cite{qameetsml,mllargespaces,smartcomp2020}. There have also been works done on adapting machine learning techniques to better guarantee robustness~\cite{marta,qvml}.

\noindent \textbf{What lies ahead:}  As depicted in Figure \ref{fig:saml_modern}, when we build an ML-based software system, there is a constant interaction between the ML process and the traditional software components. Even though different techniques for self-adaptive architecting have been proposed in the literature, ML-based software systems bring in more challenges as the uncertainties may arise not only from the software components but also from the ML components. These give rise to the following areas, which we believe requires more attention from the ML and software architecture research community as well as the practitioners in the near future:

\smallskip
\noindent \textit{1. Architectural Adaptation by ML:} This concerns the use of ML techniques for enacting architectural adaptations. For instance, the ML algorithms from the ML subsystem (Figure \ref{fig:saml_modern}) can be leveraged to predict the uncertainties that the software subsystem may encounter (e.g., overutilization of CPU, failures, etc.). Further, it can be used to autonomously select the best strategy to adapt the architecture. Such strategies might involve changing the structural composition of the software subsystem, reconfiguring the behavior of the components, etc. However, incorrect predictions of uncertainties or incorrect strategy selection can lead to sub-optimal or infeasible adaptations that may implicitly affect a system's execution. Moreover, these techniques need to take into consideration the probabilistic nature of ML processes, which might itself be the cause of uncertainty.

\smallskip
\noindent \textit{2. Architectural Adaptation for ML:} 
In an ML-based software system, the overall performance (response time, utilization, etc.) of the software is also impacted by the choice and implementation of the ML algorithm. This is because the selection of the algorithm decides how complex the final model will be. For instance, referring back to the ML subsystem in Figure \ref{fig:saml_modern}, we can make use of different algorithms to solve the same problem, which of course can offer different accuracy measures and produce different kind of computationally intensive models. A complex deep learning algorithm might produce a heavier and highly accurate model (e.g., a model with more neural network layers). But since the model is dense with more number parameters, software component(s) from the software subsystem (Figure \ref{fig:saml_modern}), leveraging the model may consume more CPU resources and takes more time to process. On the contrary, using a lighter model may offer less accuracy but may higher performance guarantees to the components consuming it (e.g., less latency for generating a prediction and thereby better response time). At runtime, over a period of time, the ML subsystem with automated MLOps practices generates different versions and types of ML models. This would require the software subsystem to make use of mechanisms to adapt autonomously between the choice of the ML models by considering the overall performance requirements of the system.

\smallskip
\noindent \textit{3. Architectural Adaptation of ML:} ML suffers from the problems of learning bias and accuracy~\cite{mlbias}. Moreover, ML models over a period suffer from model degradation where the accuracy starts reducing over time due to the variations in data. For instance, the ML subsystem (Figure \ref{fig:saml_modern}) might be making use of ML models that work very well with specific data types. However, with time as depicted in Figure \ref{fig:saml_modern}, new data also flows into the system, and this might mean the predictions of the models also start degrading. Due to this phenomenon, the components consuming the model in the software subsystem might start behaving erroneously (e.g., a recommendation system might start producing absurd recommendations to the user). This might require adapting the architecture/behavior of the neural network/ML algorithm itself to ensure robustness (e.g., dynamically adapt the layers of a deep neural network, adapt hyperparameters, etc.). This will become more relevant in the coming years, especially with the increasing popularity of Software 2.0~\cite{software2}, where software components are expected to be designed and implemented autonomously using ML techniques. 


%

Developing state of the art techniques for handling self-adaptive concerns of ML-based software systems will become more critical in the coming years. This is especially true considering the amount of run-time uncertainty that can be induced within the system due to the very nature of the ML process itself. It is also true that one cannot expect a software architect to consider all the different uncertainties at design-time. This implies that better techniques will be required to:

\smallskip
\noindent i) accurately identify uncertainties of an ML-based software system with provable guarantees.

\smallskip
\noindent  ii) autonomously select ML models from a set of available models to better guarantee the overall performance of the system without compromising much on the accuracy of the ML model.

\smallskip
\noindent iii) handle the adaptation of the ML process itself (at the algorithmic level) to better handle issues associated with robustness, fairness, etc., which may, in turn, affect the functionality of the overall system. 
    
\smallskip
\noindent \textbf{Experience:} Within the Uffizi project, we experienced both case 1 and case 2 above. In the former case, we adapted the system hardware architecture by opening or closing kiosks at run-time, depending on the actual needs forecasted by the ML model (closing kiosks was seen as an operational value since they were temporarily placed in a space that is traditionally reserved to other activities). In the latter case, since the ML model used to allocate large (competing) groups to time slots was quite complex to compute, we had to impose computational constraints on our software components to avoid response delays.

\subsection{Architecture Evolution}
Software architecture evolution refers to the process of maintaining and evolving software architectures over time based on the changes in requirements and environments. In other words, it allows the architecture to easily accommodate future changes. Evolvability of software is considered as a fundamental characteristic towards making strategic decisions and for increasing the economic value of a software~\cite{systematicevolution}.

\smallskip
\noindent \textbf{What exists:} Much work has been done in the literature that deals with software architecture evolution~\cite{systematicevolution}. Modern software systems, especially ML-based software systems, have a continuously evolving architecture. The architecture of these systems is expected to evolve at run-time continuously~\cite{woods,visionevolution,bewarefse}. This is primarily because more data becomes available as time progresses, along with the availability of newer and better algorithms. This process may call for an update of the learning algorithm. Further, this may impact changing the database schema (to incorporate new data) or adding a new database (just for storing the new data), modifying software components (due to change in the interface of the ML component), and so on. 
Further, how much data is enough and how much learning is enough at each step of the evolution requires better techniques to investigate.  \textit{Just like sometimes software architecture is driven by technological constraints, the architecture of an ML-based software system can be driven by algorithmic and data constraints}. In addition, disruptive events might completely change the system's dynamics and require a downgrade or upgrade of learned knowledge, or a previously learned knowledge might become useless. How such situations need to be handled and how the functional and non-functional requirements can be satisfied by overcoming such events is something that needs wider research and discussion. E.g., COVID and how it can impact people's behavior in entering the museum.


\smallskip
\noindent \textbf{What lies ahead:} The primary drivers on one side of ML-based systems are data and model, which is in turn based on the choice of the learning algorithm. Over a period of time, any changes in requirement arising due to the data or algorithm may require the architecture of the entire system to be evolved to satisfy the overall requirements. This brings up two essential areas that need to be considered:

\smallskip
\noindent \textit{1. Data-Induced Evolution:} One of the fundamental drivers of machine learning is the availability of good quality training data. ML-based software systems are usually implemented with good data collection pipelines responsible for collecting, filtering, and transforming data as required by the ML algorithms. The ML components further rely on the data collected to make predictions, classifications, etc. Over time, due to additional requirements, the ML algorithm may need to be adapted, and this may cause the data pipeline to be adapted to satisfy the data requirements of the ML components. This means the architect should be able to foresee such drastic changes while considering the architecture runway, which further calls for mechanisms/tools that can allow the architect to better evaluate data quality standards.

\smallskip
\noindent \textit{2. ML Algorithm-Induced Evolution:} The second fundamental driver of any ML-based software system is the choice of the learning algorithm(s) to be used and the type of model(s) generated from the learning process. Once the architecture is implemented, the ML part of the system usually keeps generating new ML models periodically based on the availability of qualitative and quantitative data. These models are further deployed into production using some standard MLOps practices. However, over time the learning algorithm might need to be modified (to produce lighter models for performance or more accurate models). This may require new data to be collected and the model interface to be redefined. This may further require the architecture of the software to be evolved to better support the new learning algorithm's requirements. This kind of evolution may also arise in the case of legacy software systems that are in need of building intelligence to existing functionalities (e.g., adding recommendation feature to traditional e-commerce system). 

One of the challenges in architecting ML-based software systems, or any software systems in general, is equipping the systems with the ability to evolve. As we have pointed above, in ML-based systems, this evolution (apart from the software side) can primarily arise from two important drivers, which include data and algorithm. However, more research needs to be done on:

\smallskip
\noindent i) how such evolution can be managed or how to better architect ML-based systems that can cope with the fluctuating demands of data and constantly evolving ML techniques.

\smallskip
\noindent ii) how to evolve a legacy software system to ML-based software systems. In such evolution, how to better handle existing team dynamics, technology constraints, organizational constraints, business requirements, etc.

\smallskip
\noindent iii) the tools/frameworks that architects can leverage to better consider an architecture runway that supports the constantly evolving requirements of an ML-based software system.

\smallskip
\noindent \textbf{Experience:} Within the Uffizi project, we built a sophisticated and precise ML model, with a 0.02 error rate, after collecting two years of data and incrementally generating a museum-specific ML model. This model was in operation during the free Sundays, for more than 20 months, till February 2020, when museums had to close due to the COVID pandemic. Now that museums are partially re-opening, and since the COVID has distorted all the rules (e.g., we have no clue on the average visiting time of tourists and how many visitors will come), can we still trust the previous data and ML model? Probably not! For instance, the COVID protocols might require the visitors to maintain a minimum distance and move in a particular manner. This might, in turn, impact the visiting time and thereby requiring additional data on visitor flow to be collected (using multiple cameras), henceforth resulting in an update to the ML model. Further, the software component(s) consuming the ML model may have to be reconfigured to accommodate the changes in the prediction accuracy/confidence score of the new ML model.



    

\section{Conclusion}
In this work, we have discussed the current outlook of software architecture in the context of ML-based systems. Based on the context of discussions put forward in light of our experience in architecting an ML-based system, we believe that future software architects' role will change from just being yet another software architect to an ML-aware software architect. In effect, there won't be a separation boundary in the vision of the architect (as depicted in Figure \ref{fig:saml_modern}).  It will be more of a unified view where there won't be any difference between software components and machine learning components at the architectural level. Rather, they will be considered as another component such as a model generator, model consumer, data/event generator, data/event consumer, etc. Such a view can then facilitate the architect to consider an ML-based software system through the lens of both machine learning and software architecture resulting in better architecting practices. In this work, we have proposed different directions that shall allow the community to propel towards such a standard. 

This is an ongoing work. The four different areas that we have considered in this paper are based on our experience while architecting the ML-based software system for queue management in Uffizi. There are certainly more areas of software architecture, such as technical debt, architecture quality, etc., which we have not covered in this work. As future work, we plan to investigate deeper into these different areas and the other research directions presented in this work. We also believe that these research directions may open up avenues for developing new architectural style(s) for ML-based software systems. However, we believe that it is a long-term process, and unlike any disruptive trend, this, too, shall pose new challenges to the software architect to wear yet another hat of a machine learning expert.


\section*{Acknowledgment}
The authors would like to thank the Uffizi Gallery for the professional support they got, especially from Director Eike Schmidt and his collaborators, Simona Pasquinucci, Gianluca Ciccardi, Serena De Laurentiis, Claudia Gerola, Angela Rossi, and Alessandra Vergari. Further, the authors would also like to thank their collaborators, Alessandro Attanasio, Maurizio Maravalle, Fabrizio Rossi, Gianluca Scatena, Francesco Tarquini, Giorgio Lattanzi, Giulio Nazzicone, Mahyar Moghaddam and Mirco Franzago from the University of L'Aquila and team Nexpecto for their constant efforts during the design, implementation, and deployment of the system.

\bibliographystyle{IEEEtran}

\bibliography{biblio}

\end{document}